\documentclass{article}
\usepackage{ijcai23}

\usepackage{times}
\usepackage{soul}
\usepackage{url}
\usepackage[hidelinks]{hyperref}
\usepackage[utf8]{inputenc}
\usepackage[small]{caption}
\usepackage{graphicx}
\usepackage{amsmath}
\usepackage{amsthm}
\usepackage{booktabs}
\usepackage[switch]{lineno}

\usepackage{stmaryrd} 
\usepackage{mathtools}

\usepackage[linesnumbered,ruled,vlined]{algorithm2e}
\usepackage{algpseudocode}
\DontPrintSemicolon

\SetKwComment{Comment}{}{}

\newcommand{\FuncCall}[2]{\texttt{\bfseries #1(#2)}}
\SetKwProg{Function}{function}{}{}

\usepackage[utf8]{inputenc}
\usepackage[T1]{fontenc}
\usepackage{zlmtt}
\usepackage{fancyvrb}

\usepackage{newfloat}
\usepackage{listings}
\DeclareCaptionStyle{ruled}{labelfont=normalfont,labelsep=colon,strut=off} 

\usepackage{multirow}
\usepackage{multicol}
\usepackage{makecell}

\usepackage{cleveref}

\newcommand{\name}{\texttt{POLIS}}
\usepackage{xcolor}

\newcommand*\GOR{\ |\ }

\newcommand{\mathbbm}[1]{\text{\usefont{U}{bbm}{m}{n}#1}}

\title{Can You Improve My Code? \\ Optimizing Programs with Local Search}

\author{
Fatemeh Abdollahi\and
Saqib Ameen\and
Matthew E. Taylor\And
Levi H. S. Lelis
\affiliations
Department of Computing Science, University of Alberta, Canada\\
Alberta Machine Intelligence Institute (Amii)\\
\emails
\{fabdolla, saqib.ameen, matthew.e.taylor, levi.lelis\}@ualberta.ca
}

\begin{document}

\maketitle

\begin{abstract}
This paper introduces a local search method for improving an existing program with respect to a measurable objective. Program Optimization with Locally Improving Search (\name) exploits the structure of a program, defined by its lines. \name\ improves a single line of the program while keeping the remaining lines fixed, using existing brute-force synthesis algorithms, and continues iterating until it is unable to improve the program's performance. \name\ was evaluated with a 27-person user study, where participants wrote programs attempting to maximize the score of two single-agent games: Lunar Lander and Highway. \name\ was able to substantially improve the participants' programs with respect to the game scores. A proof-of-concept demonstration on existing Stack Overflow code measures applicability in real-world problems.
These results suggest that \name\ could  be used as a helpful programming assistant for programming problems with measurable objectives. 
\end{abstract}

\section{Introduction}

Recent advances in large language models and program synthesis have enabled the development of powerful artificial intelligence assistants for computer programmers. For example,  Copilot~\cite{copilot} can provide an initial solution to a problem if the programmer is unsure of how to approach the problem or auto-complete what the programmer writes to speed up coding. 
Copilot and other assistants were designed to interact with the programmer throughout the development of the program. This paper considers a setting where the assistant interacts with the programmer only \emph{after} a working version of the program is available. In this paper's setting, the assistant attempts to improve the programmer's solution with respect to a real-valued, measurable objective function, something systems such as Copilot cannot perform.  


We introduce Program Optimization with Locally Improving Search (\name), an intelligent assistant to improve existing programs. \name\ leverages the ability of existing synthesizers to generate high-quality (short) programs by treating each line of an existing program as an independent program synthesis task. \name\ uses an enumeration algorithm for synthesis, called bottom-up search~\cite{AlbarghouthiGK13,Udupa:2013}, for each line of the program. Since \name\ selects the best solution encountered in each bottom-up search, it can be seen as a hill-climbing algorithm in the program-line space. Despite not using any models for guiding its search, \name\ can handle complex programs because it divides the original problem into much smaller sub-problems by considering the synthesis of one line at a time. 

To evaluate \name, 27 programmers wrote programs for playing Lunar Lander and Highway, two single-agent games commonly used to evaluate reinforcement learning algorithms. \name\ was able to improve the score of \emph{all} programs written by the participants, often by a large margin. Our results also show that often the modified programs retain most of the structure of the original programs. As a result, the users who wrote the programs are likely to understand \name's modifications to their implementations. We also present a proof-of-concept demonstration of \name's ability of fixing bugs in 4 simple programs posted on Stack Overflow. 

\name's modified programs can be seen as the result of the work done by an effective human-AI team. This is because bottom-up search would not be able to synthesize the resulting programs from scratch, as the programs are  long and complex. However, bottom-up search is able to substantially improve human-generated programs. As our results demonstrate, human programmers are unable to write on their own programs of the quality obtained with \name. These results suggest that \name\ can be a helpful assistant to programmers for problems with measurable objectives. 

This paper makes two contributions. First, it defines a problem setting for intelligent programming assistants where the assistant attempts to improve existing programs with respect to an objective function. Second, it introduces \name, a system that employs a novel local search algorithm based on a simple brute-force search algorithm. 

\section{Related Work}

\name\ is related to intelligent programming assistants, program synthesis, programmatically interpretable policies, and program enhancement algorithms. 

\subsection{Intelligent Programming Assistants}

Intelligent assistants for programmers are getting popular and have become a popular area of research lately. SnipPy~\cite{small-step-live} is one such tool that allows the programmer to synthesize instructions by defining input-output examples in the context of live programming. Similarly, Blue-Pencil~\cite{on-the-fly-edits} is a system that identifies repetitive tasks that arise in programming and suggests transformations for such tasks. reCode~\cite{recode} observes code transformation to identify other places of the code that would require similar changes. \citeauthor{code-completion-statistical}~\shortcite{code-completion-statistical} introduced a statistical model for code completion and \citeauthor{guo2022learning}~\shortcite{guo2022learning} introduced a model for code completion that leaves ``holes'' where the model is uncertain.  

\name\ differs from these works in \emph{how} it assists the programmer. Instead of real-time interactions during the development of the program, we consider the scenario where the programmer provides a complete, compilable version of their program. \name\ leverages human-defined code structure to improve the user's implementation with a simple synthesizer. 

\subsection{Program Synthesis}

The task of synthesizing programs that satisfy a specification is a long-standing problem~\cite{1969:PROW,1976:Smith,1985:AutoSynthesis,1990:LogicProgrammingAndSynthesis} and it has received much attention lately~\cite{AlbarghouthiGK13,BalogGBNT16,robustfill17,kalyan2018neuralguided,Shin2019SyntheticDF,dreamcoder}. While previous works attempt to improve the synthesis process and generate programs which satisfy given specification, \name\ uses program synthesis to optimize existing programs with respect to a given objective function.

\subsection{Programmatic Policies}

One way to solve the problems considered in this work is to synthesize programs encoding a policy for solving the tasks. Neurally directed program search (NDPS) \cite{NDPS} synthesizes programs while imitating a neural oracle. Viper~\cite{viper} also employs imitation learning to train decision trees encoding policies. In order to provide better search guidance for synthesis, Propel~\cite{propel} trains neural policies that are not ``too different'' from the synthesized program. Sketch-SA~\cite{sketch-sa} is another such system that uses imitation learning to synthesize a sketch of a policy; the policy is synthesized from the sketch by evaluating it directly in the environment. 

Oracle-free programmatically interpretable reinforcement learning ($\pi$-PRL) \cite{qiu2022programmatic} and Bilevel Synthesis (Bi-S)~\cite{bilevel} bypass the need of an oracle to guide the synthesis of programmatic policies. $\pi$-PRL uses a differentiable language and trains the model using policy gradient methods, while Bi-S uses the result of a search in a feature space to guide the search in the programmatic space.

\name\ differs from these algorithms because they were designed to synthesize programs from scratch, while \name\ focuses on leveraging the structure of existing programs. 

\subsection{Program Enhancement}

Refactoring is a well-known program enhancement technique used to improve a program's quality without affecting its external behavior \cite{fowler2018refactoring,ml-refactoring}. Another way of enhancing a program is the Automated Program Repair (APR) technique which refers to the process of fault localization in software and the development of patches using search-based software engineering and logic rules~\cite{survey-code-analysis,automated-program-repair,semfix}. For instance, \citeauthor{1genprog}~\shortcite{1genprog} use genetic programming to develop bug-fixing patches without affecting software functionality. \name\ is different from these techniques because a) \name\ improves programs with respect to an objective function and its external behavior is likely to change; and b) while \name\ fixes unintended programmer mistakes (similar to APR), it is likely to also change  sub-optimal parts of the program, improving overall performance.

\section{Problem Definition}

Rather than using a general-purpose language like Python, which defines a very large program space, we use a domain-specific language (DSL) to define a more constrained space of programs for solving a programming task. A DSL is defined as a context-free grammar $(V, \Sigma, R, S)$, where $V$ is a finite set of non-terminals, $\Sigma$ is a finite set of terminals, and $R$ is the set of relations corresponding to the 
production rules of grammar. $S$ is the grammar's start symbol. An example of a DSL defined by a grammar $G$ is shown below, where $V = \{S, C, B\}$, $\Sigma = \{c_1, c_2, c_3, b_1, b_2$, if-then-else$\}$, $R$ are the relations (e.g., $C \to c_1$), and $S$ is the start symbol. 
\begin{align}
\label{eq:DSL}
S &\to \text{if}(B) \text{ then } S \text{ else } S \nonumber \GOR C  \\ 
C &\to c_1 \GOR c_2 \GOR c_3 \GOR CC \\
B &\to b_1 \GOR b_2 \nonumber 
\end{align}

This DSL allows programs with a single instruction ($c_1, c_2,$ or $c_3$), or multiple commands using nested if-then-else blocks. Let $\llbracket G \rrbracket$ be the set of programs (possibly infinite) that can be written with grammar $G$. Each program $p \in \llbracket G \rrbracket$ is defined by a pair $\{T, L\}$, where $T$ is a multiset of non-terminal symbols and $L$ defines a partition of symbols from $T$ into program lines, i.e., $L$ defines how a programmer organizes the symbols in $T$ in a text editor. Note that two programs that have identical functionality could have different partitions $L$. 

\name\ takes as input a program $p \in \llbracket G \rrbracket$, and an objective function $F$ (real-valued evaluation of the program), and outputs a program $p' \in \llbracket G \rrbracket$ that is at least as good as $p$ and approximates a solution for $\arg\max_{p \in \llbracket G \rrbracket} F(p)$, assuming a maximization problem. 

\section{\name: A Programming Assistant}

\begin{algorithm}[t]
    \caption{\name}
    \label{algo:localsearch}
    \KwData{Initial program $p$, overall time limit $t$, time limit per line $t_{l}$, evaluation function $F$}
    \KwResult{Improved program $p'$}
    \While{\FuncCall{Not-Timeout}{$t$}}{
        $p' \gets p$\;
        \For{$i \gets 1$ to \FuncCall{Number-of-Lines}{$p$}}{ \label{algo:for_loop}
          $p \gets$ \FuncCall{synthesizer}{$p, i, t_{l}, F$}\;\label{algo:probecall} 
        }
        \Comment{\footnotesize{\# search has reached a local minimum}}
        \If{\FuncCall{F}{$p$} = \FuncCall{F}{$p'$}}{return $p'$}\label{algo:break}
    }
    return $p'$
\end{algorithm}



The pseudocode in Algorithm~\ref{algo:localsearch} shows the local search algorithm \name\ employs. It receives an existing program $p$ and two time limits, $t$ and $t_l$, for the overall running time of the search and for the running time allowed to optimize each line of code, respectively, and an evaluation function $F$. \name\ returns a new program, $p'$, that is at least as good as $p$ in terms of $F$-value. While there is time available to improve the input program, \name\ iterates through each line (the for loop in line~\ref{algo:for_loop}) and it attempts to synthesize a program that replaces the code in the $i$-th line of $p$ such that the objective function $F$ is improved. This is achieved with a call to the synthesizer (line~\ref{algo:probecall}), which returns a version of $p$ where the $i$-th line of $p$ is replaced by a program that optimizes $F$. The synthesizer can return the program unchanged, if its original $i$-th line returns the best $F$-value or it exceeds its time limit before finding a better line. Lastly, \name\ returns the optimized program (line \ref{algo:break}) if the search reaches a local optimum, i.e., the improved program $p$ has the same $F$-value as $p'$.

Our system uses size-based bottom-up search (BUS)~\cite{AlbarghouthiGK13,Udupa:2013} as the synthesizer. BUS was shown to outperform other uninformed enumeration-based synthesizers \cite{probe}. BUS starts by enumerating the smallest possible programs of a given language. It then uses the smallest programs with the production rules of the DSL to generate larger programs. One can use different metrics of ``size'' for defining BUS's enumeration procedure. A commonly used metric, which we use in our implementation, is the number of nodes in the abstract syntax tree representing the synthesized programs. That is, in BUS's first iteration it generates all programs whose tree has a single node, then all programs whose tree has two nodes, and so on, until a solution is found. In its first iteration, for the DSL shown in Equation~\ref{eq:DSL}, BUS generates programs $c_1, c_2, c_3, b_1, b_2$. Then, in its second iteration BUS generates programs $c_1 c_1$, $c_1 c_2$, $c_1 c_3$ $c_2 c_2$, $c_2 c_1$, $c_2 c_3$, and so on. One advantage of BUS is that, once it finds a solution program, the program is provably the smallest one that solves the problem. Another advantage is that all programs generated in search are executable, which allows one to run them and perform an observational equivalence check (i.e., the search only keeps one of two programs that produce the same set of output values for a given set of input values of interest). 

\subsection{Domain-Dependent Implementation Details}

We evaluate \name\ on programmatic policies for playing games, which are written by human programmers. A programmatic policy is a program encoding a function (policy) that receives a state of a game and returns the action the agent should take at that state. In what follows, we describe \name's implementation details. 


\subsubsection{Input-Output Examples}


For the task of writing programmatic policies for playing games, we use the approach introduced by \citeauthor{pirl}~\shortcite{pirl} to define a set of input-output examples. That is, we train a neural policy that generates a set of input-output pairs: for a set of observations $o$ (input), we store the neural policy's chosen action $a$ (output). We use DQN~\cite{mnih2015human} to train a neural policy $\pi$ for 2000 episodes. We let the agent follow $\pi$ in the environment for 2000 steps and collect all the observation-action pairs along with their $Q$-values.


\subsubsection{Evaluation Function}

We use two evaluation functions. The function $F$ is given by running the programmatic policy and computing its game score. This evaluation function is computationally expensive, since we need to play the game several times to evaluate a program, due to the stochastic nature of the environments. 
Instead of computing $F$ for all programs generated in search, we keep a list of the current $k$-best programs with respect to an action-agreement metric: the number of observations each program correctly maps to the action a neural policy $\pi$ selects for that observation. The action-agreement metric we use is computed as $\frac{\sum_{o \in T} \mathbbm{1}[p(o) = \pi(o)]}{|T|}$, where $T$ is the set of input-output examples, $\mathbbm{1}[\cdot]$ is the indicator function, $p(o)$ and $\pi(o)$ are the actions returned by the program $p$ and policy $\pi$, respectively, for observation $o$.  We evaluate the value of $F$ only for the programs in the $k$-best set. 
Once the synthesizer runs out of time, it returns the best program in the set of $k$ best with respect to $F$, not with respect to the action agreement metric. We use $k=20$ in our experiments. 





\subsubsection{Highlights} 

Highlights ranks a set of observations according to the largest difference in $Q$-values for different actions available at a given observation. We employ the idea of highlights to further optimize the computational cost of our evaluation function by using a small number of input-output examples. Instead of collecting a large number of observation-action pairs uniformly at random, we collect the 400 observations ranked most important by Highlights~\cite{highlights}. 

\subsubsection{Bayesian Optimization} 

The real numbers $n$ in the DSL (Figure~\ref{fig:dsl_experiments}) are set using Bayesian optimization~\cite{bayesian_optimization}. Bottom-up enumeration in the synthesizer generates programs with the symbol $n$, later replaced with real values by the optimizer. The optimizer chooses these values while attempting to optimize for the action agreement metric. 


\subsubsection{Restarts}

The initial program and the set of input-output pairs define the optimization landscape \name\ traverses with its hill-climbing algorithm. \name{}'s greedy approach to optimization could lead to the algorithm returning locally optimal solutions. An effective strategy for dealing with local optimum solutions is to restart the search from a different starting location in the optimization landscape once the search stops in a local optimum~\cite{local_search}. To restart the search and allow for different initial starting conditions, we train a different DQN agent to generate a new set of input-output pairs every time we restart the algorithm. A restart is triggered in Algorithm~\ref{algo:localsearch} when line~\ref{algo:break} is reached and \name\ still has time available for synthesis. 

\section{User Study Evaluation}
This section describes the experimental design of the study.\footnote{Our \name\ implementation and the data collected in our user study is available at \url{https://github.com/FatemehAB/POLIS}.} 

\begin{figure}[!t]
    \centering
    \includegraphics[width=150px]{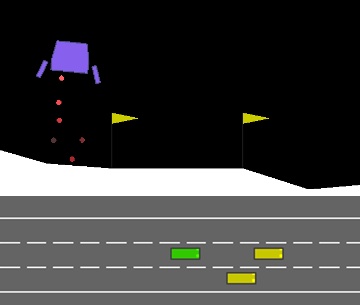}
    \caption{Lunar lander (top) and Highway (bottom)}
    \label{fig:games}
\end{figure}

\subsection{Problem Domains}

We use \name\ to improve programs written by users to play two games: Lunar Lander and Highway (Figure~\ref{fig:games}). 
Both games have a game score, which serves as a clear metric for evaluating the quality of the programs. 

\paragraph{Lunar Lander} In this game the player controls three thrusters of a spaceship trying to land on the moon. Each thruster can be either on or off. The game score is maximized if the player does not use the thrusters unnecessarily and gently reaches the landing pad. We use the LunarLander-v2 implementation from OpenAI Gym~\cite{openai-gym}. 

\paragraph{Highway} In this game the player controls a car on a three-lane highway. The game score is higher when the player drives fast, avoids collisions, and spends more time in the rightmost lane. The player can change lanes, increase, or reduce speed. We use the implementation of \citeauthor{highway-env}~\shortcite{highway-env}.

\subsection{User Study Design}

\begin{figure}[!t]
\begin{align*}
P &\Coloneqq \textbf{def } \text{heuristic(o): S } \textbf{return } \text{action}\\
S &\Coloneqq SS \GOR \textbf{if}\text{(C): S }\textbf{else}\text{: S} \GOR \textbf{if}\text{(C): S }\textbf{elif}\text{(C): S }\textbf{else}\text{: S} \\
 & \ \ \ \ \ \ \GOR V_{\text{def}} \GOR A_{\text{assign}} \\
C &\Coloneqq \text{C B C} \GOR \text{E} \\
E &\Coloneqq o_{i} \GOR V_{\text{name}} \GOR n \GOR \text{E M E} \GOR \text{pow(E, E)} \GOR \text{sqrt(E)}  \\
&  \ \ \ \ \ \ \GOR \text{log(E)} \GOR \text{-(E)} \GOR \text{abs(E)}\\
V_{\text{def}} &\Coloneqq V_{\text{name}}\text{ = E}\\
A_{\text{assign}} &\Coloneqq \text{action = }a_{i}\\
B &\Coloneqq \text{and} \GOR \text{or} \GOR < \GOR <= \GOR != \GOR == \GOR >= \GOR > \\
M &\Coloneqq + \GOR - \GOR * \GOR /
\end{align*}
\caption{DSL used by \name\ for games domain; $o$ is the set of observations passed as input to the program and $a_{i}$ and $o_{i}$ refer to one of the actions and observations of the game; $n$ is a real number. $V_{\text{def}}$ refers to the declaration of a variable with name $V_{\text{name}}$ and $A_{\text{assign}}$ to the assignment of an action to the agent.}
\label{fig:dsl_experiments}
\end{figure}

We developed a web-based system based on HIPPO Gym~\cite{hippogym} to conduct the user study and advertised it in mailing lists of graduate and undergraduate Computing Science students at our university.\footnote{The study was approved by the University of Alberta Research Ethics Office (Pro00113586).} Each participant first electronically signed a consent form, explaining that they would write a program to play a computer game. It also explained that their compensation would be impacted by the game score of their final program; higher game scores would result in higher monetary compensation. The minimum compensation was \$15. We used the following formulae to compute the compensation of each participant: $15+ (100+x) \times (1/30)$ and $15 + x \times (1/5)$ for Lunar Lander and Highway, respectively. $x$ represents the participants' average game score over 100 and 25 episodes of Lunar Lander and Highway, respectively (an episode is completed when the player finishes landing the spaceship in Lunar Lander or when the player crashes the car or a time limit is reached in Highway). The maximum compensation was capped at $\$25$.

After agreeing with the terms of the study, each participant was randomly assigned to one of the two games. Then, they read a tutorial about the assigned game. In the tutorial, we explained the features in each observation passed as an input parameter to the program as well as the actions available to the player. Our tutorial had a few examples with screenshots of the game showing situations where different actions were applied to different observations of the game. The tutorial finished with a multiple-choice question about the game; immediate feedback was provided to the participant showing whether they chose the correct or wrong answer. If an answer was incorrect, the participant would have as many attempts as needed to answer it correctly. 

Following the game tutorial, each participant read a tutorial about our DSL. The tutorial presented the DSL (Figure~\ref{fig:dsl_experiments}) and explained Boolean and algebraic expressions as well as the programming structures our DSL supports. Similarly to the game tutorial, we provided several examples of programs that can be written in our DSL. The tutorial finished with a multiple-choice question where the participant had to select, among four options, the program that was accepted in our DSL; the participant had as many attempts as needed to answer the question correctly.  

Before writing a program for playing the game, the participant had the chance to play the game using their keyboard for a maximum of 10 minutes. Our graphical user interface showed, in real-time, the observation values and the game score each participant obtained for each run of the game. The participant could choose to stop playing the game at any time (within the 10 minutes allowed by our system) and start writing their program. Our goal with this step of the study was to allow the participant to develop a strategy for playing the game, something they could try to encode in their programs.

We provided the participants with a Python-like editor, where the keywords of the DSL are highlighted. The editor also had an example of a simple program for playing the game. For Highway, the initial program moves the car to the right lane if the car is not already there; the player takes no action otherwise. Our interface also allowed the participants to go back to the tutorials while writing their program. 

Our interface also showed the game so that participants could execute their program and see its behavior. Similarly to the interface where the participant played the game, we showed the observation values and the game scores in real-time. The participant could stop the simulation at any time to inspect the values of the observations. We stored all programs the participants evaluated so that they could be used as input for our evaluation. The total time allowed for the experiment was 60 minutes. The participant could submit the final version of their program at any time within the 60-minute limit. We used the final program submitted to compute the participant's monetary compensation. The participant then answered demographic questions before leaving.

\section{User Study Results}



In our results, we abbreviate standard deviation as SD and interquartile range as IQR.

\subsection{Demographics}
40 people consented to participate and 26 completed the survey. The average age was 20.96 (SD of 4.13), with their ages ranging from 18 to 40; 20 of the participants identified themselves as male, 5 as female, and 1 withheld gender information.
Most (20) had received or were pursuing undergraduate education, 4 had completed high school, and 2 were pursuing post-secondary training. Most (25) had not done any form of game artificial intelligence (AI) research and about half of them had not taken any AI courses. More than one-third of the participants (10) rarely or never played computer games and others occasionally or often played computer games. 

We asked about the participants' programming experience: 22 had more than one year of experience and 4 had less than a year. We also asked about their knowledge of Python, how hard it was to write a program in our DSL, and how hard it was to write a program for solving the game. We used a 5-point, Likert-like scale: 1 being ``novice'' in Python and ``very easy'' for writing programs, to 5 being ``expert'' in Python and ``very hard'' for writing programs. The median response to these three questions were: 3 (IQR = 1), 2.5 (IQR = 2), and 4 (IQR = 1), respectively. On average, the participants had some experience in Python, and found it easy to use our DSL, but found it hard to write a program to play the game. To evaluate \name\, we considered the data from those who submitted at least one working program (different from the example program we provided), resulting in a total of 27 participants (one of them did not complete the survey). 

\begin{figure}[h!]
\begin{lstlisting}
def heuristic(o):
    action = 0
    if (o[5] == o[1] and o[5]-o[1] > 200) or (o[9] == o[1] and o[9]-o[1] > 200):
        action = 4
    elif (o[5] == o[1] and o[5]-o[1] <= 200) or (o[9] == o[1] and o[9]-o[1] <= 200):
        if o[1] == 4:
            if o[9] < 4:
                action = 2
            else:
                action = 0
        else:
           action = 0
    else:
        action = 3
    return action
\end{lstlisting}
\caption{Example of a program written for the Highway domain by a participant of the user study.}
\label{fig:example_program}
\end{figure}

\subsection{Computational Results}

\Cref{ltable,htable} show the results for Lunar Lander and Highway, respectively. Here, each participant is represented by an ID. The game score of both the participants' and \name{}'s programs is an average of the score the program obtained in 100 of Lunar Lander and 25 episodes of Highway. The game score shown for \name\ is the average over 10 independent runs of the system. Each run of \name\ can result in different game scores due to the random initialization of the neural policy used to generate input-output pairs. We also present the standard deviation, minimum, and maximum game scores across these 10 independent runs. We performed 5 restarts for each run of the system; the result of a run is the best program encountered across the 5 restarts. The average score we present for both participants and \name\ are for the program that achieved the highest average score throughout the study; the program the participant submits is not necessarily the program with the highest score. The number of lines of code (LoC) indicates how many lines the original program has. In both tables, we sort the rows according to the participant's program game score, from lowest (top) to highest (bottom). The number of edited lines (Edited LoC) refers to the average number of lines that \name\ modifies in the restart that resulted in the best program of a given run. We also show the average number of car collisions in Highway (Hits). 

\name{}'s average score is higher for all programs written in our study. Even the minimum value across the 10 independent runs is often much higher than the score of the program the participants wrote. A Wilcoxon signed-rank test pointed to a large effect size for the average results of both domains: $0.624$ for Lunar Lander ($p<4.9 \times 10^{-4}$) and $0.621$ for Highway ($p < 3.1 \times 10^{-5}$).

For Lunar Lander, \name\ provided quite significant improvements to some of the participants' scores (e.g., IDs 3 and 11), but for some others the improvements were minor (e.g., IDs 4 and 5). The number of lines edited for the programs of participants 4 and 5 is much smaller than for the other programs, which indicates that \name\ quickly reached a local minimum for these programs. Interestingly, for Highway, \name\ improved the performance of all programs to an average game score above $33$ (the best program a participant wrote achieved a score of $35.71$). Moreover, \name\ substantially reduced the number of collisions, in some cases from more than 20 to less than 3 collisions. Since \name\ does not change the overall structure of the program, we conjecture that the participants identified the program structure needed to play Highway, which makes the programs for that game more amenable to \name{}'s improvements. 
The Lunar Lander results might be pointing to a limitation of \name\, which is its inability to improve programs that need simultaneous changes to more than one line of code.  

\begin{table}[h!]
    \setlength\tabcolsep{0pt}
    \small
    \centering
    \begin{tabular*}{\columnwidth}{r @{\extracolsep{\fill}} *{7}{r} }
        \toprule
        \addlinespace[5pt]
        \multirow{2}{*}{ID} & \multicolumn{2}{c}{Original Program} & \multicolumn{5}{c}{{\name} Program}\\
        \addlinespace[3pt]
        \cmidrule(l){2-3}
        \cmidrule(l){4-8}
        {} & \multicolumn{1}{c}{Score} & \multicolumn{1}{c}{LoC} & \multicolumn{1}{c}{Score} & \multicolumn{1}{c}{SD} & \multicolumn{1}{c}{Min} & \multicolumn{1}{c}{Max} & \makecell{Edited \\ LoC} \\
        \midrule
        1 & -449.72 & 8 & -22.17 & 9.18 & -35.03 & -7.95 & 11.40\\
        2 & -221.98 & 34 & 151.40 & 24.24 & 112.12 & 190.45 & 27.30\\
        3 & -196.33 & 14 & 68.85 & 37.46 & -22.09 & 105.62 & 13.90\\
        4 & -125.12 & 9 & -123.96 & 3.45 & -125.12 & -113.61 & 1.10\\
        5 & -125.12 & 14 & -122.68 & 7.30 & -125.12 & -100.79 & 1.10\\
        6 & -118.65 & 22 & 102.31 & 25.68 & 38.67 & 134.19 & 28.80\\
        7 & -110.72 & 29 & 0.44 & 35.71 & -34.20 & 90.06 & 9.80\\
        8 & -87.42 & 16 & 56.19 & 21.45 & 23.15 & 96.78 & 17.70\\
        9 & -79.13 & 26 & 96.04 & 24.00 & 63.09 & 141.24 & 29.60\\
        10 & -70.13 & 10 & 65.05 & 17.06 & 38.68 & 101.32 & 5.90\\
        11 & -21.50 & 52 & 252.90 & 6.60 & 240.77 & 260.86 & 30.50\\
        12 & 52.84 & 13 & 74.45 & 7.99 & 63.84 & 89.30 & 5.30\\
        \bottomrule
    \end{tabular*}
    \caption{Game score improvements for Lunar Lander}
    \label{ltable}
\end{table}

\begin{table}[h!]
    \setlength\tabcolsep{0pt}
    \small
    \centering
    \begin{tabular*}{\columnwidth}{r @{\extracolsep{\fill}} *{9}{r} }
        \toprule
        \addlinespace[5pt]
        \multirow{2}{*}{ID} & \multicolumn{3}{c}{Original Program} & \multicolumn{5}{c}{{\name} Program}\\
        \addlinespace[3pt]
        \cmidrule(l){2-4}
        \cmidrule(l){5-10}
        {} & \multicolumn{1}{c}{Score} & \multicolumn{1}{c}{LoC} & \multicolumn{1}{c}{Hits} & \multicolumn{1}{c}{Score} & \multicolumn{1}{c}{SD} & \multicolumn{1}{c}{Min} & \multicolumn{1}{c}{Max} & \makecell{Edited \\ LoC} & \multicolumn{1}{c}{Hits} \\
        \midrule
        1 & 5.09 & 40 & 25 & 35.72 & 0.36 & 34.64 & 35.84 & 7.90 & 4.70\\
        2 & 6.21 & 8 & 25 & 33.44 & 2.12 & 30.25 & 35.84 & 5.90 & 2.90\\
        3 & 6.21 & 6 & 25 & 35.24 & 0.60 & 34.64 & 35.84 & 4.00 & 3.50\\
        4 & 7.25 & 24 & 25 & 35.46 & 0.86 & 34.64 & 36.59 & 5.50 & 3.20\\
        5 & 9.72 & 11 & 24 & 36.29 & 0.45 & 35.84 & 36.74 & 3.00 & 4.50\\
        6 & 10.74 & 6 & 24 & 35.12 & 0.59 & 34.64 & 35.84 & 5.20 & 3.20\\
        7 & 11.36 & 20 & 24 & 36.09 & 0.77 & 34.64 & 37.43 & 8.90 & 3.30\\
        8 & 12.74 & 8 & 23 & 34.88 & 0.48 & 34.64 & 35.84 & 5.20 & 2.60\\
        9 & 14.23 & 16 & 23 & 35.21 & 0.74 & 34.64 & 36.74 & 4.30 & 3.10\\
        10 & 15.33 & 9 & 22 & 35.00 & 0.55 & 34.64 & 35.84 & 5.30 & 2.90\\
        11 & 16.17 & 12 & 23 & 34.98 & 1.77 & 30.25 & 36.74 & 3.70 & 3.50\\
        12 & 28.98 & 10 & 11 & 36.01 & 1.11 & 34.64 & 37.74 & 3.50 & 3.50\\
        13 & 30.98 & 14 & 8 & 35.35 & 0.86 & 34.07 & 36.74 & 3.30 & 4.50\\
        14 & 31.95 & 22 & 2 & 40.20 & 2.29 & 37.43 & 42.25 & 4.10 & 0.10\\
        15 & 35.71 & 13 & 2 & 37.08 & 0.15 & 37.02 & 37.53 & 2.00 & 1.00\\
        \bottomrule
    \end{tabular*}
    \caption{Game score improvements for Highway}
    \label{htable}
\end{table}

\subsection{Representative Program}

The program shown in Figure~\ref{fig:example_program} is a representative program written by one of the participants of our study for the Highway domain; we refer to this program as $p$ in this section. This program obtains an average game score of $6.8$ over 25 episodes. Figure~\ref{fig:example_improvement} shows \name{}'s improved program for $p$, which we will refer to as $p'$. We lightly edited $p'$ for readability. 
\name{}'s $p'$ obtains an average game score of $39.0$ over 25 episodes, a major improvement over the original program. The participant of our study made a mistake while writing the first if-statement of $p$ as the Boolean condition checks whether $o[5]$ is equal to $o[1]$ and if $o[5] - o[1] > 200$; the two parts of the expression cannot be simultaneously true as once $o[5]$ is equal to $o[1]$, we have that $o[5] - o[1]$ is zero. As a result, the player never slows down (action $4$). The participant's intention with this if-statement was likely to slow the car down if the player's car was on the same lane as the nearest car on the road (the condition ``$o[5]$ is equal to $o[1]$'' returns true if the cars are on the same lane). 

\name\ not only fixed the problem with the Boolean condition in the participant's program, but also changed the player's strategy. Instead of slowing down if another car is on the same lane, $p'$ only slows down when changing lanes; $o[3]$ is the car's velocity on the $y$-axis, which is different from zero when the car is changing lanes. Since the car is changing lanes, $o[1]$ cannot be zero, as $o[1]$ is zero when the car is in the leftmost lane. Unlike $p$, $p'$ changes lanes when there is another car in the same lane. This is encoded in the elif structure of the program, which can be translated as if the nearest car is on the same lane ($o[5]$ is equal to $o[1]$) and the car is not already in the rightmost lane (line 7), then move to the right lane (action $2$; line 8). The agent will move to the left lane if already in the rightmost lane (action $0$; line 10). 

\name{}'s improved program prefers to drive in the rightmost lane if the car driving in the same lane is not the closest (i.e., there is still time to change lanes). The program maximizes its score by driving in the rightmost lane. Finally, \name{}'s program does nothing (action $1$) if it is not changing lanes and there is no car in front of it. \name{}'s strategy is a cautious one as the car slows as it changes lanes, but never accelerates. This cautious strategy achieves a much higher game score than the participant's program. 

\begin{figure}[h]
\begin{lstlisting}
def heuristic(o):
    action = 0
    if o[1] and o[3]:
        action = 4
    elif o[5] == o[1] or o[9] == o[1]:
        if o[1] == o[5]:
            if o[1] < 7.9317:
                action = 2
            else:
                action = 0
        else:
           action = 2
    else:
        action = 1
    return action
\end{lstlisting}
\caption{\name's improved program for the program written by a participant in user study (Figure~\ref{fig:example_program}).}
\label{fig:example_improvement}
\end{figure}

\section{Proof of Concept: Stack Overflow}

\begin{figure}[!t]
\begin{align*}
P &\Coloneqq \textbf{def } \texttt{optimized(*args)} : S \textbf{ return } *args\\
S &\Coloneqq SS \GOR \textbf{while}(C): S  \GOR \textbf{if}(C): S \\
 & \ \ \ \ \ \ \GOR \textbf{if}(C): S \ \textbf{else}: S \GOR \textbf{if}(C): S \ \textbf{elif} \ (C): S \\
& \ \ \ \ \ \ \GOR  \textbf{if}(C): S \ \textbf{elif} \ (C): S \ \textbf{else}:S \\
 & \ \ \ \ \ \ \GOR V_{\text{def}} \GOR Arr_{\text{assign}} \\
C &\Coloneqq \text{C B C} \GOR \text{E} \\
E &\Coloneqq Arr[E] \GOR V_{\text{name}} \GOR n \GOR \text{E M E} \\
 & \ \ \ \ \ \ \GOR \text{pow(E, E)} \GOR \text{sqrt(E)} \GOR \text{log(E)} \GOR \text{-(E)} \GOR \text{abs(E)}\\
V_{\text{def}} &\Coloneqq V_{\text{name}}\text{ = E}\\
Arr_{\text{assign}} &\Coloneqq Arr[E] = E\\
B &\Coloneqq \text{and} \GOR \text{or} \GOR < \GOR <= \GOR != \GOR == \GOR >= \GOR > \\
M &\Coloneqq + \GOR - \GOR * \GOR /
\end{align*}
\caption{DSL used by \name\ for Stack Overflow problems.}
\label{fig:SO_dsl_experiments}
\end{figure}

To demonstrate that \name\ is general and can be applied to problems other than games and also to languages with more complex structures such as loops, we collected four programs with implementation problems on Stack Overflow and translated them to our Python-like language so that \name could fix them. Three of the four programs are sorting algorithms; the last program attempts to compute the cumulative sum of a set of numbers. 
Figure~\ref{fig:SO_dsl_experiments} shows the DSL used in this experiment. The input parameter \texttt{*args} indicates that the program \texttt{optimized} can accept and return a variable number of arguments depending on the problem being solved. Compared to the DSL used in the user study, this DSL accepts more data types (arrays) and more complex structures (loops). 

\name\ corrected all three sorting programs with the evaluation function that simply counts the number of input examples that are correctly mapped to the desired output. The problems with the Stack Overflow sorting programs were simple (e.g., one of the programs used \texttt{<} instead of \texttt{<=} in the Boolean expression of a while loop) and \name\ was able to fix them by changing a single line of the original programs.

The fourth program we collected on Stack Overflow attempts to solve a ``cumulative sum problem,'' which is defined as follows. Given an array of numbers, the goal is to replace each element with index $i$ in the array with the sum of all elements with index $j \leq i$. For example, the expected output for array $[4,3,6]$ is $[4, 7, 13]$. Figure~\ref{fig:sum} shows the incorrect implementation of a program for solving the cumulative sum problem (\texttt{sum\_wrong}) and \name's corrected version for the problem (\texttt{sum\_fixed}). The cumulative sum program had two implementation errors: the Boolean expression of the while-loop and the list used in the operation within the loop. \name\ could not fix them by simply using the number of input examples correctly mapped to the desired outputs. Instead, we used an $F$ function that computed the sum of the absolute differences between each element of the list the program produced as output and the desired list of numbers. Using this $F$ function, \name\ corrected the program, as shown in Figure~\ref{fig:sum}.

In this proof-of-concept experiment, we manually generated the input-output examples, similar to how a programmer would come up with a set of test cases for their program. Such a set could possibly be used to define \name's $F$ function, so it can attempt to correct the implementation errors.

\begin{figure}
\begin{lstlisting}
def sum_wrong(array, n, sums):
    i = 0
    while i < n-1:
        sums[i] = array[i] + array[i+1]
        i = i+1
    return sums

def sum_fixed(array, n, sums):
    i = 0
    while i < n:
       sums[i] = array[i] + sums[i - 1]
       i = i + 1
    return sums
\end{lstlisting}
\caption{Example of an incorrect program posted on Stack Overflow, where }
\label{fig:sum}
\end{figure}


\section{Conclusions}

In this paper, we present \name, a system capable of improving existing programs with respect to a measurable, real-valued metric. \name\ employs a simple synthesizer within the loop of its local search. \name\ divides the problem of improving an existing implementation into smaller sub-problems by considering each line of the program as an independent program synthesis task. This way, \name\ employs a bottom-up search synthesizer that attempts to replace a single line of the original program at a given time, while all the other lines remain unchanged. We conducted a user study where 27 participants wrote programs to play two games. \name\ was able to improve the performance of the programs of all participants, often by a large margin. Since \name\ performs local changes with an enumerative synthesizer, its modified program shares the same structure as the original program. The similarity of the programs allowed us to understand how \name\ was able to improve the performance of a representative program from our study. We also performed a proof-of-concept experiment with four programs collected from Stack Overflow to demonstrate that \name\ can also be applied to other application domains and handle more complex languages such as those with loops. \name\ was able to correct all four programs. The results of our experiments suggest that \name\ can be used as a programming assistant in scenarios where one is interested in improving an existing program with respect to a measurable, real-valued metric.  

\section*{Acknowledgements} This research was supported by Canada's NSERC and the CIFAR AI Chairs program. The research was carried out using computational resources from Compute Canada. 
Part of this work has taken place in the Intelligent Robot Learning (IRL) Lab at the University of Alberta, which is supported in part by research grants from the Alberta Machine Intelligence Institute (Amii); a Canada CIFAR AI Chair, Amii; Compute Canada; Huawei; Mitacs; and NSERC. 
We thank the anonymous reviewers for their feedback. 

\bibliographystyle{named}
\bibliography{ijcai23}

\end{document}